# Integer and Fractional Quantum Hall effect in Ultra-high Quality Few-layer Black Phosphorus Transistors

Jiawei Yang[1]*, Son Tran[1]*, Jason Wu[2], Shi Che[1], Petr Stepanov[1], Takashi Taniguchi[3], Kenji Watanabe[3], Hongwoo Baek[4], Dmitry Smirnov[4], Ruoyu Chen[1†], Chun Ning Lau[1†]

[1] Department of Physics, Ohio State University, Columbus, OH 43220
[2] Department of Physics and Astronomy, University of California, Riverside, CA 92521
[3] National Institute for Materials Science, 1-1 Namiki Tsukuba Ibaraki 305-0044 Japan.
[4] National High Magnetic Field Laboratory, Tallahassee, FL 32310
* These authors contribute equally to this work.

Abstract
As a high mobility two-dimensional semiconductor with strong structural and electronic anisotropy, atomically thin black phosphorus (BP) provides a new playground for investigating the quantum Hall (QH) effect, including outstanding questions such as the functional dependence of Landau level (LL) gaps on magnetic field $B$, and possible anisotropic fractional QH states. Using encapsulating few-layer BP transistors with mobility up to 55,000 cm$^2$/Vs, we extract LL gaps over an exceptionally wide range of $B$ for QH states at filling factors $\nu$=-1 to -4, which are determined to be linear in $B$, thus resolving a controversy raised by its anisotropy. Furthermore, a fractional QH state at $\nu \sim$ -4/3 and an additional feature at -0.56 ± 0.1 are observed, underscoring BP as a tunable 2D platform for exploring electron interactions.

The quantum Hall (QH) effect is a prototypical 2D phenomenon[1]. In a perpendicular magnetic field $B$, the cyclotron orbits of charge carriers coalesce to form discrete Landau levels (LLs), which, combined with confining potentials, give rise to topologically protected edge modes that suppress backscattering and support dissipationless transport. While integer QH states may arise from either single particle effect (such as cyclotron or Zeeman gaps) or symmetry-breaking processes due to electronic interactions, fractional QH effect is a manifestation of strong electronic interactions that are only observed in very high $B$ and/or ultrahigh quality devices. Despite intense investigation over the past three decades, several outstanding questions remain unanswered, such as the anisotropy of fractional states in higher LLs[2], and the possibly non-Abelian nature of the even denominator states that cannot be accounted for by the otherwise successful composite fermion theory[3-14].

The advent of 2D materials, whose atomically thin structures afford stronger electronic interactions and more competing symmetries, has renewed interest in this celebrated phenomenon. In particular, integer QH effect has been observed in monolayer[15, 16] and few-layer graphene[17, 18], transition metal dichalcogenides[19, 20] and InSe[21], though graphene remains the only 2D material in which fractional QH states are observed[22-29].

---

[†] Emails: chen.7729@osu.edu; lau.232@osu.edu

As a relatively new member of the family of 2D materials[30-32], few-layer BP boasts the highest hole mobility in 2D semiconductors[32-36]. It provides a platform to study fundamental phenomena such as the quantum Hall effect[35, 36] or topological transitions[37-39], with potential electronic, thermal and optoelectronic applications. Integer quantum Hall effect has been observed in both electron[40] and hole-doped regimes[33, 35, 36, 41], though LL gaps at low filling factors are measured only at very high magnetic fields and over limited range (27 T ≤ B ≤ 33 T).

A rather unusual property of BP is its strong structural anisotropy, where the unequal nearest neighbor hopping rates along different directions result in a highly asymmetric band structure. Consequently, its dispersion is nearly linear along the *x*-direction, thus is expected to be Dirac-like, but quadratic or Schrödinger-like along the *y*-direction[42] (Fig. 1a), giving rise to strong anisotropy in electronic properties[31, 43, 44]. At finite $B$, the cyclotron orbits of charge carriers are elliptical rather than circular. Though the LL gaps $\Delta$ are not expected to exhibit directional dependence, the anisotropy *could* manifest in the anomalous dependence on $B$. For instance, in a system with linear-quadratic hybrid dispersions, instead of the $(NB)^{1/2}$-dependence in the Dirac limit and $(N+1/2)B$-dependence in conventional semiconductors, the LL energies are expected to scale as $((N+1/2)B)^{2/3}$ [45], where $N$ is the LL index. In the case of BP, such sub-linear Landau spectrum might be expected owing to the mixture of Dirac-like band structure into BP[45, 46], though other theories argue that such sub-linear behavior only emerge under large strains or electric field[37, 39, 47-49]. The experimental challenge in determining the functional dependence of LL gaps lies in fabricating devices with *demonstrated anisotropy* as well as sufficiently high quality that enable measurement of Landau level gaps over an extended range of magnetic fields.

In this Letter, using ultrahigh quality BP devices with field effect mobility up to 55,000 cm$^2$/Vs and strong anisotropy in conductivity, we report observations of the IQHE at magnetic fields as low as 10 T, and determination of LL gap scaling for QH states at filling factors -1 ≤ υ ≤ -4. The LL gaps are predominantly linear in B, despite the observed anisotropic conductivity. At very high magnetic fields, we observe fractional QH states at filling factor υ~-4/3 and υ~-0.56 ± 0.1. As the first observation of FQHE in a non-graphene 2D material, our work shed light on electron/hole correlations and providing a new playground for exploring FQHE states with possible even-denominator states.

Ultrahigh quality BP devices are fabricated by encapsulating BP sheets between hexagonal boron nitride (hBN) layers. Bulk hBN and BP crystals are synthesized via high temperature and high-pressure techniques. Their thin flakes are exfoliated onto Si/SiO$_2$ substrates, and a dry transfer technique is applied to assemble hBN/BP/hBN stacks[50]. Both the exfoliation and transfer steps are completed inside a VTI glove box, with moisture and oxygen concentration lower than 0.1 ppm. After encapsulation, the stacks are taken out of the glove box and etched with SF$_6$ plasma into either Hall bar or van der Pauw geometry (Fig. 1b). The etching directions are chosen to align with the long straight edges of the BP sheets, so that they are likely along the crystallographic directions of the lattice. A second etching step is applied to remove the top hBN and expose the BP layers, followed by immediate e-beam evaporation of Cr/Au metal electrodes, resulting in high quality Ohmic contacts. The devices are characterized at room temperature and low temperature using standard dc and lock-in techniques.

The BP devices consistently have high quality, as attested by Fig. 1c, which displays four-terminal resistance $R_{xx}$ as a function of $V_{bg}$. At low temperature, $\mu_{FE}$ up to 55,000 cm$^2$/Vs is observed (Fig. 1c). At room temperature, we observe a record high field effect mobility $\mu_{FE}$ ~ 2000 cm$^2$/Vs, exceeding previous reports by a factor of 2-10 (Fig. 1c inset).

The electronic anisotropy of the devices are investigated using the van der Pauw geometry. The longitudinal and transverse resistances are measured by $R_{xx}=V_{12}/I_{43}$, $R_{yy}=V_{23}/I_{14}$, and $R_{xy}=V_{13}/I_{24}$, respectively, where $V_{ij}$ ($I_{ij}$) denote bias voltage (measured current) between leads $i$ and $j$ (Fig. 1b). Fig. 1d displays the transfer curves of $R_{xx}$ and $R_{yy}$ measured along two perpendicular directions as a function of back gate voltage $V_{bg}$ at 300 mK. The charge neutrality point (CNP) is at $V_{bg}$ = 0.4 V, suggesting that the pristine BP flake is intrinsic. We observe a high field effect hole mobility of 6,000 cm$^2$/Vs along the $x$ direction, and a much suppressed but still respectable mobility of 1,100 cm$^2$/Vs along the $y$ direction. The overall conductivity along the $x$ direction is also much higher. As $R_{xx}$ and $R_{yy}$ are measured using the same contacts, the measured anisotropy in mobility and conductivity cannot result from different contact resistances, but is intrinsic to the device. Such strong electronic anisotropy is reproduced in three high mobility devices, in agreement with prediction from band structure calculations[51-56] and prior experiments[43].

In a high magnetic field, when the Fermi level is pinned between the discrete LLs, the device exhibits vanishing longitudinal resistance $R_{xx}$ and Hall resistance $R_{xy}$ quantized at $R_Q/\nu$, where $R_Q=h/e^2$ is the resistance quantum and $\nu=nh/Be$ the filling factor. Here $e$ is the electron charge, $h$ the Planck constant, and $n$ the charge carrier density. Fig. 2a displays the Landau fan diagrams $R_{xx}(n, B)$ for $5 \leq B \leq 15$T. Shubnikov–de Haas (SdH) oscillations or integer QHE are well-resolved over the entire range. As the formation of LLs averages over all orientations, the Landau fan diagrams of $R_{xx}$ and $R_{yy}$ exhibit no qualitative difference, as expected, though at low field the QH features do resolve better along the $x$ direction. The quantum Hall states at $\nu$ = -3 and -4 are resolved at as low as $B$ = 6 T, and $\nu$ = -1 at $B$ =9 T. We note that this is the lowest field at which QH plateaus are resolved, indicating the highest quantum mobility to date. Fig. 2b displays the experimental signature of the IQHE at B =18 T, where $R_{xx}$ vanishes while $R_{xy}$ plateaus at quantized values. Fig. 2c displays Hall resistance $R_{xy}$ as a function of gate voltage $V_g$ at $B$=18, 20, 22, 26, 28 and 30 T, respectively. Well-quantized plateaus are observed; when plotted against $\nu=nh/Be$, the data collapse into a single curve (Fig.2d) with well-quantized plateaus at $R_Q/\nu$ for $\nu$=-1, -2, -3, -4, -5 and -6, respectively. The well-resolved odd states suggest a full lifting of the spin degeneracy.

To determine the scaling of LL gaps with $B$, we measure the minima of longitudinal resistance at constant filling factors and magnetic field as a function of temperature, and repeat the measurements at different $B$ (Fig. 3a). In Arrhenius plots, the data points at different filling factors approximately fall on a straight line (Fig. 3b), and can be satisfactorily described by the thermally activation model,

$$R_{xx,\ min} = Ae^{-\Delta_\nu/2k_BT} \qquad (1)$$

where $A$ is a constant pre-factor, $\Delta_\nu$ the LL gap for the QH state at filling factor $\nu$ and $k_B$ the

Boltzmann's constant. Fitting the data points to Eq. (1) yield $\Delta_\nu$ at a given magnetic field and filling factor. Due to the high Landau level degeneracy at high magnetic fields and the limited range of gate voltage, only the lowest 6 LLs are accessible in this device.

We first focus on odd filling factors, at which the LL gaps are expected to arise solely from Zeeman splitting $E_Z = g_s\mu_B B$, thus scale linearly with $B$. Here the measurement of LL gaps provides a way to determine the Landé factor and offers insight into the energetics of Landau levels. Indeed, for $\nu=-1$ and $-3$, plotting the experimentally measured gaps versus $B$ yield straight lines (Fig. 3c-d), which are fitted to the equation

$$\Delta_{odd}= g_s\mu_B B - 2\Gamma_\nu \quad (2)$$

where $g_s$ is the effective Landé factor that can be enhanced from its bare electron value $g_0=2$ through exchange interactions, $\mu_B$ is Bohr magneton and $\Gamma_\nu$ represents the LL broadening. Fitting the data points to Eq. (2) yield $g_s=$ 4.4, 4.3 and 2.1 at $\nu=-1$, $-3$ and $-5$ respectively. The enhancement of $g_s$ at $\nu=-1$ and $-3$ originates from the exchange interactions between charge carriers with net spin polarization. Thus, the Landé factor is expected to oscillate between the odd and even states, and adopts the maximum values at odd states. At $\nu=-5$, the population of additional LLs weakens the net spin polarization, and the Landé factor recovers the free electron value. The Landau level broadening $\Gamma_\nu$ is extracted from the negative intercepts at $B=0$, which is estimated to be ~ 10 K for all three states at $\nu=-1$, $-3$ and $-5$, indicating the lowest disorder in BP samples to date.

For QH states at even filling factors, the LL gaps consist of contributions from both the cyclotron gap $E_c$ and the Zeeman energy $E_Z$,

$$\Delta_{even}= E_c - g_0\mu_B B - 2\Gamma_\nu \quad (3)$$

Here Landé $g$-factor assumes the bare value of $g_0=2$, due to the absence of spin polarization of the highest filled LL. For conventional semiconductors, the Landau level is expected to scale linearly with B, as $E_c= \hbar eB/m^*$ where $m^*$ is the effective mass of charge carriers. However, due to BP's anisotropic crystal and electronic structures[44], the scaling of its cyclotron gap in $B$ has been under considerable theoretical debate[46-49].

We can now directly address this outstanding controversy, enabled by the ultra-high mobility of our devices and the large range of magnetic field over which QH plateaus are resolved. Our measurements of $\Delta_{-2}$ and $\Delta_{-4}$ are presented in Fig. 3e-f, where reasonable linearity as a function of magnetic field has been observed. Fitting the data points to Eq. (3) and assuming the conventional expression for $E_c$, we extract $m^*\sim 0.33 m_e$, where $m_e$ is the rest mass of electrons, in agreement with prior works[33-36, 57, 58]. $\Gamma_\nu$ is estimated to be ~10 K, consistent with those found for odd integer states. The confidence of the fits is ~ 92%. Thus we conclude that, within ~8% error, the cyclotron gaps in BP are linear in $B$. The deviation of our results from the sublinear behavior predicted in ref 45 may arise from the presence of a band gap in pristine BP, instead of the gapless graphene model. Thus, the anomalous $B^{2/3}$ scaling of Landau level gaps could manifest under large strain in BP, which is predicted to reduce and even close the band gap[37].

One of the foremost goals of investigation of BP in the QH regime is the QHE states, as the comparable energy scales of $E_c$ and $E_z$ are expected to facilitate the observation of even–denominator states[59, 60], thus providing another much-needed playground to study this very intriguing state for topological computing. Moreover, BP's anisotropic crystal lattice may shed light on the anisotropic FQHE states observed in GaAs systems[3-14]. Indeed, at very high magnetic fields, we observe a FQHE state in device 2. Fig. 4a plots the $R_{xx}$ and $R_{xy}$ line traces vs $V_{bg}$ at B = 45 T. A well-defined minimum in $R_{xx}$ appears at $V_{bg}$~ -25 V (indicated by black arrows), accompanied by a plateau in $R_{xy}$ quantized to ~19 kΩ ($3/4\ h/e^2$). Both features move with $n$ and $B$, as indicated by the dashed lines in Fig. 4b, and disappear at $B$~ 41 T (Fig. 4b). We thus attribute this feature to a fraction QH state. For better visualization, we replot the same data in Fig. 4c, where red, blue and orange data points represent the center of the $\nu$=-1, -2 and -3 QH plateaus, respectively, and the black the observed fractional state. The dashed lines extrapolated to $B$=0, where they converge at the band edge. From their slopes, the filling factor of the fractional state is estimated to be -4/3, in agreement with the plateau value $R_{xy}$.

Additionally, in device 1 with van der Pauw geometry, we have observed another feature that occurs at a fractional filling factor. Fig. 4d plots differentiated $dR_{xx}/dB$ as a function of $n$ and $B$ for 25≤$B$≤45T. In addition to the well-resolved integer states, a kink appears above $\nu$=-1 for $B$>35T. The kink moves with both $n$ and $B$, i.e. at a constant filling factor, indicating its origin as a QH state. Fig. 4e displays line traces of $R_{xx}$ as a function of $V_{bg}$ at $B$=35, 37, 38, 41, 43, 44 and 44.5 T, respectively, with black arrows indicating the feature that moves with $B$, with a constant filling factor is estimated to be $\nu$~ -0.56±0.1. Due to the limited resolution of the data, the exact fraction of the state is difficult to pinpoint. Considering that the 1/3-demoninator states typically have the largest LL gap, we tentatively attribute the observed feature to $\nu$=−2/3 state, though more experimental studies of higher mobility samples are necessary to determine the nature of this state.

Both fraction QH states are smeared out at temperature ~2-3 K; considering that the LL broadening in these devices is 2$\Gamma$~ 20 K, the corresponding fractional gaps is ~25 K at $B$=45 T. This value is larger than or comparable to those in GaAs heterostructures with much higher mobility[61-63]. With further optimization of device quality, we expect BP to provide a platform for exploring novel FQHE such as anisotropic states and even-denominator states that may be tunable by strain and electric field.

In conclusion, we successfully fabricated few-layer BP transistors with ultrahigh quality. All integer QH states are resolved at relatively low magnetic fields. A systematic study of the LL gaps over a wide range of magnetic field for the first four LLs reveals a linear magnetic field scaling, thus resolved an outstanding debate that originate from the presumed linear-quadratic hybrid dispersion of BP. We also report the first observation of fractional QH states in few-layer BP systems in multiple devices, paving the way to the future study of correlated phenomena in this new playground.

**Acknowledgement**
This work is supported by NSF/ECCS 1509958. A portion of this work was performed at the National High Magnetic Field Laboratory, which is supported by National Science Foundation Cooperative Agreement No. DMR-1157490 and the State of Florida. K.W. and T.T. acknowledge support from the Elemental Strategy Initiative conducted by the MEXT, Japan and JSPS KAKENHI Grant Numbers JP26248061, JP15K21722 and JP25106006.

Fig. 1. Band structure, device schematics and transport characteristics at $B=0$. (a). Band structure of few-layer black phosphorus near the $\Gamma$ point. (b). Schematics of van der Pauw devices. Inset: typical van der Pauw device. (c). $R_{xx}(V_{bg})$ at $T=4K$ of a high quality Hall bar device, with field-effect mobility ~55,000 $cm^2$/Vs. Inset: $R_{xx}(V_{bg})$ in same units as main panel at 300K, with mobility ~ 2000 $cm^2$/Vs. (d). Anisotropic longitudinal resistances $R_{xx}$ and $R_{xy}$ vs $V_{bg}$ for a van der Pauw device at $T=4K$.

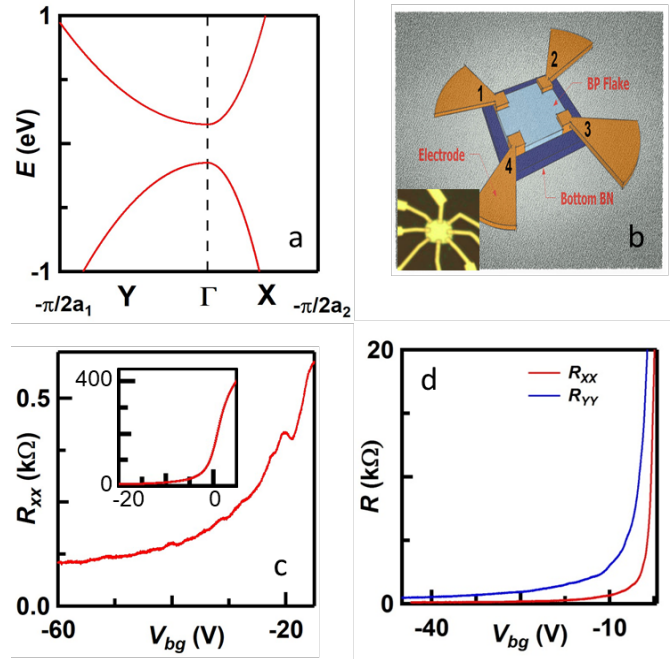

Fig. 2. Magnetotransport data. (a). $R_{xx}(V_{bg}, B)$ at $T=0.3$K. (b). $R_{xx}(V_{bg})$ and $R_{xy}(V_{bg})$ line traces at $B=18$ T. (c). $R_{xy}(V_{bg})$ line traces at different magnetic fields, which collapse onto a single curve $R_{xx}(\nu)$ in (d).

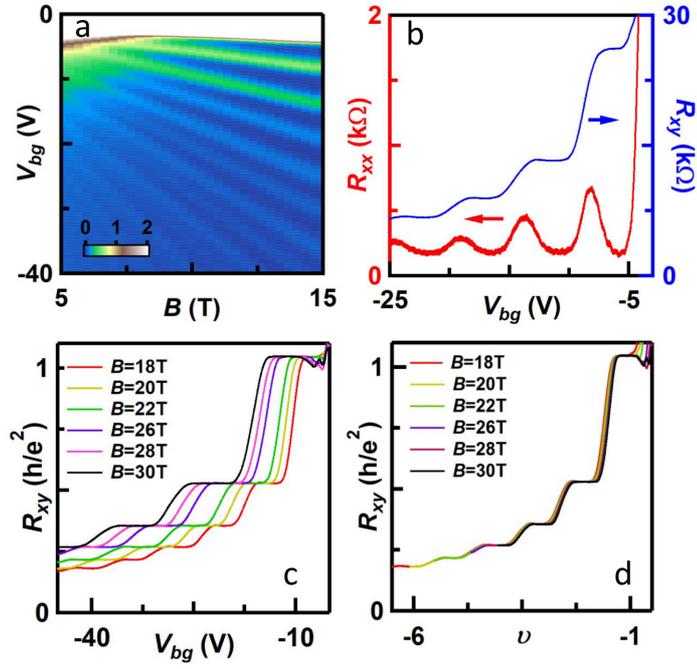

Fig. 3. Extraction of LL gaps. (a). $R_{xx}(V_{bg})$ at temperatures from 0.9K to 14.6K at $B$=18 T. (b). Arrhenius plot of $R_{xx}$ minima vs $1/T$ at different filling factors at $B$=18 T. (c-f). Extracted LL gaps $\Delta(B)$ for $v$=-1, -2, -3 and -4, respectively.

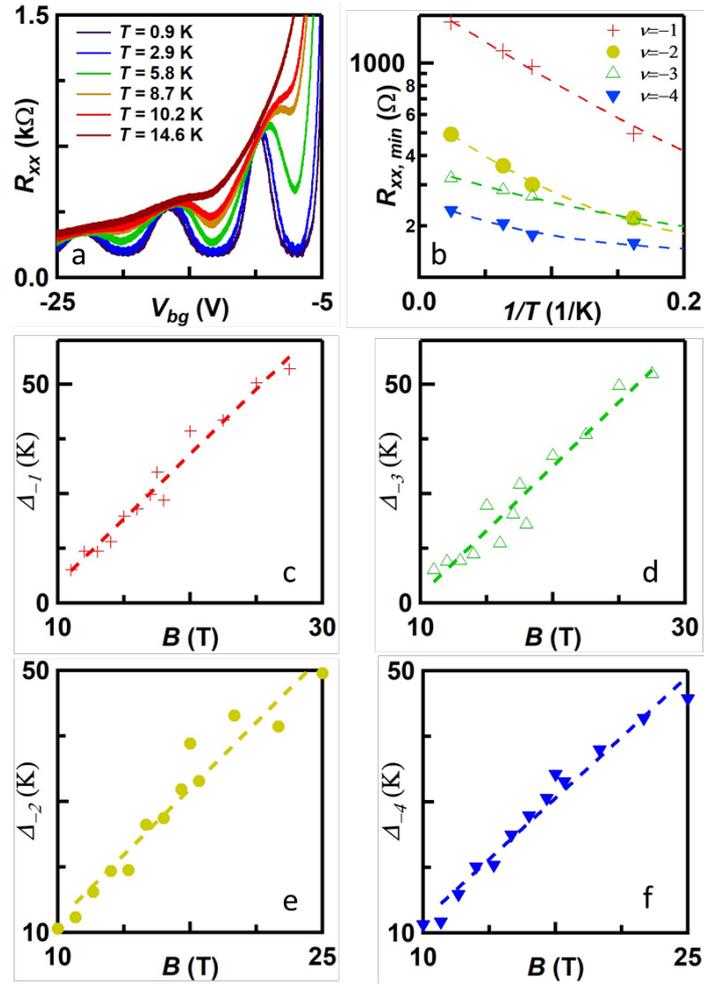

Fig. 4. Fractional QH states in a Hall bar (a-c) and van der pauw device (d-e). (a). $R_{xx}$ (left axis) and $R_{xy}$ (right axis) of a Hall bar device vs $V_{bg}$ at $B=45$ T. The fraction QH state manifests as a small plateau in $R_{xy}$ and an accompanying dip in $R_{xx}$, as indicated by the black arrows. (b). $R_{xx}(V_{bg})$ at different magnetic fields, showing the fractional feature moving with field. Lines are offset for clarity. (c). A constructed Landau fan diagram from line traces shown in (b), converging to the band edge at $B=0$. (d). A Landau fan diagram $dR_{xx}/dB$ $(V_{bg}, B)$ of a van der Pauw device. Numbers denote the integer filling factors. The fractional feature above $\nu=-1$ is indicated by the arrow. (e). Line traces of $R_{xx}(V_{bg})$ at $B$ from 36T to 44.5 T (blue to red) at intervals of 1T.

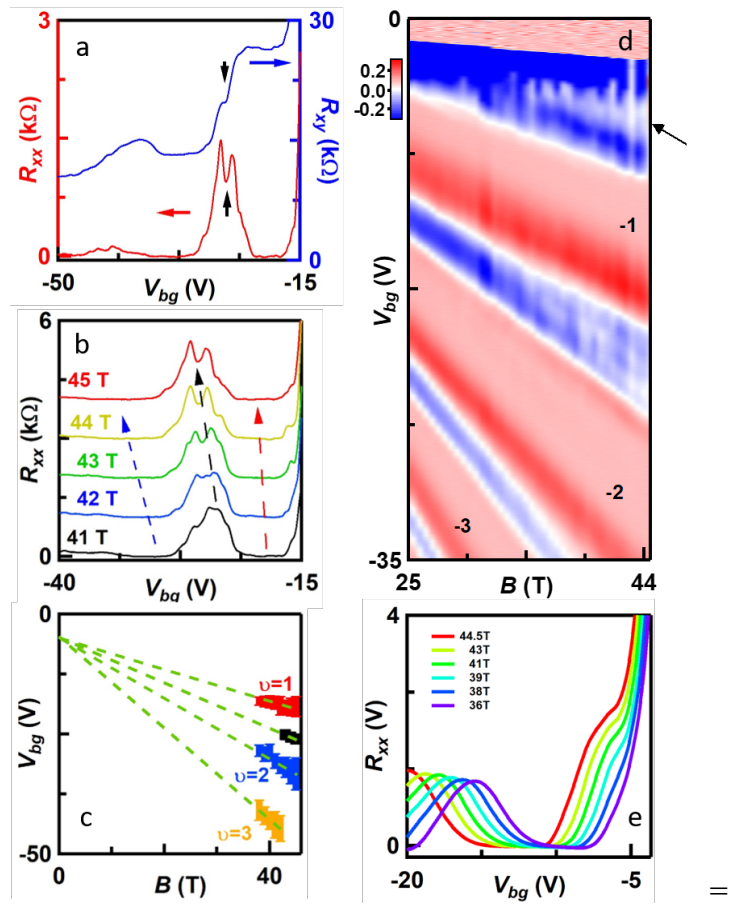